\documentclass[sigconf, nonacm]{acmart}

\renewcommand{\paragraph}[1]{\noindent\textbf{#1.}}

\usepackage{xcolor}
\definecolor{R1color}{RGB}{0,102,204} 
\usepackage{enumitem}
\setlist[itemize]{labelsep=3pt,leftmargin=*}
\setlist[enumerate]{labelsep=3pt,leftmargin=*}

\usepackage[dvipsnames]{xcolor}
\definecolor{nrblue}{RGB}{0,146,245}

\usepackage{subcaption}

\usepackage{xspace}
\usepackage{tabularx}
\usepackage{array}

\usepackage{listings}
\definecolor{PigBlue}{RGB}{42, 0, 255}
\definecolor{PigRed}{RGB}{255, 0, 0}
\usepackage{todonotes}

\newcommand{\sysname}{{NeurEngine}}

\newcommand{\aidb}{AI$\times$DB\xspace}

\usepackage{lipsum}

\usepackage[most]{tcolorbox}
\newtcolorbox{myquote}[1]{
    enhanced, 
    colback=white, 
    colbacktitle=nrblue, 
    colframe=nrblue,
    sharp corners, 
    left=2pt,
    right=2pt,
    top=2pt,
    bottom=2pt,
    title={\textcolor{white}{#1}},
    boxrule=1pt,
}

\usepackage{marginnote}
\definecolor{metacolor}{HTML}{0072B2} 
\definecolor{R1color}{HTML}{800080} 
\definecolor{R2color}{HTML}{008866} 
\definecolor{R3color}{HTML}{D55E00} 

\newtcolorbox{myquoteMeta}[1][]{
    colback=black!5,
    colframe=black!5,
    notitle,
    sharp corners,
    borderline west={1.5pt}{0pt}{blue},
    enhanced,
    breakable,
    left=2pt,
    right=2pt,
    top=0pt,
    bottom=0pt,
    ignore nobreak,
}

\newtcolorbox{myquoteR1}[1][]{
    colback=black!5,
    colframe=black!5,
    notitle,
    sharp corners,
    borderline west={1.5pt}{0pt}{R1color},
    enhanced,
    breakable,
    left=2pt,
    right=2pt,
    top=0pt,
    bottom=0pt,
    ignore nobreak,
}

\newtcolorbox{myquoteR2}[1][]{
    colback=black!5,
    colframe=black!5,
    notitle,
    sharp corners,
    borderline west={1.5pt}{0pt}{R2color},
    enhanced,
    breakable,
    left=2pt,
    right=2pt,
    top=0pt,
    bottom=0pt,
    ignore nobreak,
    #1
}

\newtcolorbox{myquoteR3}[1][]{
    colback=black!3,
    colframe=black!3,
    notitle,
    sharp corners,
    borderline west={1.5pt}{0pt}{R3color},
    enhanced,
    breakable,
    left=2pt,
    right=2pt,
    top=0pt,
    bottom=0pt,
    ignore nobreak,
}

\usepackage{multicol}

\begin{document}

\title{Towards Effective Orchestration of AI {x} DB Workloads}

\author{Naili Xing}
\email{xingnl@comp.nus.edu.sg}
\affiliation{%
  \institution{National University of Singapore}
  \country{Singapore}
}

\author{Haotian Gao}
\email{gaohaotian@comp.nus.edu.sg}
\affiliation{%
  \institution{National University of Singapore}
  \country{Singapore}
}

\author{Zhanhao Zhao}
\authornote{Corresponding author.}
\email{zhzhao@comp.nus.edu.sg}
\affiliation{%
  \institution{National University of Singapore}
  \country{Singapore}
}

\author{Shaofeng Cai}
\email{shaofeng@comp.nus.edu.sg}
\affiliation{%
  \institution{National University of Singapore}
  \country{Singapore}
}

\author{Zhaojing Luo}
\email{zjluo@bit.edu.cn}
\affiliation{%
  \institution{Beijing Institute of Technology}
  \city{Beijing}
  \country{China}
}

\author{Yuncheng Wu}
\email{wuyuncheng@ruc.edu.cn}
\affiliation{%
  \institution{Renmin University of China}
  \city{Beijing}
  \country{China}
}

\author{Zhongle Xie}
\email{xiezl@zju.edu.cn}
\affiliation{%
  \institution{Zhejiang University}
  \city{Hangzhou}
  \country{China}
}

\author{Meihui Zhang}
\email{meihui_zhang@bit.edu.cn}
\affiliation{%
  \institution{Beijing Institute of Technology}
  \city{Beijing}
  \country{China}
}

\author{Beng Chin Ooi}
\email{ooibc@zju.edu.cn}
\affiliation{%
  \institution{Zhejiang University}
  \city{Hangzhou}
  \country{China}
}

\begin{abstract}
AI-driven analytics are increasingly crucial to data-centric decision-making.
Executing relational and AI operators in separate runtimes prevents the
database optimizer and runtime from coordinating operator ordering, model
placement, batching, and state reuse.
Integrating AI operators into database engines enables such coordination but
raises challenges in jointly optimizing query processing and model execution,
scheduling under resource contention, and reusing relational intermediates and
AI artifacts.

This paper formalizes \aidb workloads as iterative, concurrent, and shareable
executions that interleave relational and AI operators.
We then advocate \emph{database-native orchestration} as a paradigm for redesigning
database engines for these workloads and distill two design principles:
holistic \aidb co-optimization and unified \aidb cache management.
We present \sysname{} as a proof-of-concept prototype and report preliminary
results illustrating the performance benefits of \emph{database-native orchestration}.
\end{abstract}

\maketitle

\section{Introduction}
\label{sec:intro}

The rapid rise of AI agents is fundamentally transforming the execution of data-centric tasks~\cite{YaoZYDSN023,SchickDDRLHZCS23,JiangXSSZLZLLHWZYCC25,xing2026ptnas}.
Given a natural-language task description, an AI agent, typically built around an LLM, acts as an autonomous orchestrator that executes a multi-step, adaptive loop to reach a solution~\cite{HongZCZCWZWYLZR24}.
In each iteration, the agent may consult metadata to ground its plan, retrieve targeted data via relational queries, and invoke analytical models to process the retrieved content.
The agent then validates intermediate results and refines its workflow until it yields the final results~\cite{EckmannB26}.
Throughout this process, relational operators (e.g., join and aggregation) 
are interleaved with AI operators (e.g., training and inference).
We refer to these integrated, end-to-end workflows as \aidb workloads.

Despite the growing prevalence of \aidb workloads, production systems still predominantly rely on an inefficient \emph{export-execute-import} paradigm, where data is extracted from the database, processed in external runtimes, and ultimately written back~\cite{singa, JiangXSSZLZLLHWZYCC25,FathollahzadehMB25,ZhaoZL24,rafiki}.
While recent efforts have attempted to bring AI closer to data via SQL-oriented abstractions (e.g., UDFs)~\cite{XingCCLOP24,ZengXCCOPW24,FoufoulasPS25}, a critical gap remains: 
\textit{existing database engines are not natively designed to orchestrate \aidb workloads that couple fundamentally divergent execution models.}
Databases excel at global plan optimization, relational query execution over stored data, transaction management, and controlled concurrency~\cite{SelingerACLP79,HellersteinSH07,PanCDWCCO26}.
In contrast, AI runtimes primarily optimize local model execution, leaving global orchestration decisions such as operator ordering, model placement, batching, and state reuse outside the database. 
As a result, treating AI operators as black boxes obscures their cost-quality trade-offs, data dependencies, and resource footprints. 
This lack of transparency hinders co-optimization between relational and AI operators, and prevents unified management of shared intermediates, leading to suboptimal performance for AI$\times$DB workloads~\cite{FoufoulasPS25,abs-2503-23863,ZhaoGXZZCRO26}.

To bridge this gap, we envision a new class of database engines that provide
\emph{database-native orchestration} for \aidb workloads.
We first formalize this workload and distill three characteristics that distinguish it from traditional analytical workloads:
(i) \textit{iterative}, as execution proceeds through an adaptive loop of exploration and refinement;
(ii) \textit{concurrent}, as AI agents generate many parallel requests while exploring alternative solution paths or retrying failed steps; and
(iii) \textit{shareable}, as substantial overlap exists in data retrieval, intermediate computations, and AI-specific artifacts across iterations and concurrent executions.
Together, these characteristics motivate the co-design of the query optimizer and execution runtime.
We organize this co-design around two principles:
(i) holistic \aidb co-optimization, which jointly optimizes relational and AI operators under data, model, and resource constraints;
and (ii) unified \aidb cache management, which enables the reuse of relational intermediates and AI artifacts across iterations and concurrent executions.
Building on these principles, we formulate an open research agenda by specifying their objectives and identifying key challenges.
Finally, we present {\em \sysname{}} as a proof-of-concept implementation of these principles, together with preliminary results illustrating the potential benefits of database-native orchestration.

We hope this preliminary exploration encourages the data management community to rethink database design for \aidb workloads and address their emerging challenges.

\section{BACKGROUND and VISION}
\label{sec:vision}

\paragraph{\aidb Workload} 
We define \aidb workload as follows.
\begin{definition}[\aidb Workload]
An \aidb workload $\mathcal{W}=\langle q_1,\dots,q_n\rangle$ is a sequence of
\aidb queries issued over a database instance $\mathcal{D}$, often by AI agents.
Each query interleaves relational and AI operators and accesses or produces AI
artifacts such as model versions, embeddings, and model caches.
\end{definition}
We summarize key characteristics of \aidb workloads.
\begin{itemize}
    \item \textit{Iterative.} An \aidb workload follows a multi-step
    adaptive loop in which an AI agent repeatedly issues queries to probe data,
    validate results, and refine its plan until reaching a solution.
    \item \textit{Concurrent.} Within a workload, an AI agent may issue multiple
    queries in parallel to explore alternative solution paths, execute independent
    steps, or retry failures, resulting in bursty arrivals and resource contention
    at the database engine.
    \item \textit{Shareable.} Queries across iterations and concurrent executions
    often overlap in data retrieval, intermediate computations, and model
    invocations, creating potential for reuse and consolidation.
\end{itemize}
\paragraph{\aidb Query}
We define \aidb query as follows.
\begin{definition}[\aidb Query]
An \aidb query $q$ is a declarative program whose logical plan composes DB operators with AI operators from $\Omega_{DB} \cup \Omega_{AI}$.
Here, $\Omega_{DB}$ includes relational operators (e.g., $\sigma,\pi,\Join$ for selection, projection, and join) for data access, and
$\Omega_{AI}$ includes AI operators (e.g., $\texttt{PREDICT}(\cdot)$).
\end{definition}
As a concrete example, consider ad click-through-rate prediction on the multi-table RelBench Avito dataset~\cite{relbench}.
Given a natural-language request (\textit{predict each candidate ad's click-through rate and, based on the predictions over different time windows, decide which ads to promote immediately, which to run in the short-term campaigns, which to allocate the weekly budget to, and which to reduce exposure for}), the AI agent translates the request into an \aidb workload: a task graph with three horizon-specific \aidb prediction queries for 1-day, 3-day, and 7-day CTR prediction, followed by a cross-horizon aggregation task.
Each \aidb query constructs point-in-time features and invokes \texttt{PREDICT}; the aggregation task joins the predictions into actions.
Listing~\ref{lst:query} shows the 3-day query.

This workload exhibits the three properties above: it is iterative because the action-aggregation task depends on horizon-specific prediction queries, each of which may be refined through repeated invocations with different model settings;
concurrent because the 1-day, 3-day, and 7-day prediction queries can be issued in parallel; and shareable because these queries share relational data, feature-construction steps, and model invocation contexts.

\paragraph{Core Design Principles}
These characteristics reveal a mismatch between \aidb workloads and traditional DBMSs, motivating a redesign of the database engine to deliver higher efficiency for \aidb workloads.
We advocate \emph{database-native orchestration} as the paradigm for this redesign, where optimization, scheduling, and execution are coordinated inside the DBMS.
We define it as follows.
\begin{definition}[Database-Native Orchestration]
Database-native orchestration is an architectural paradigm for \aidb workloads in which AI operators and their artifacts are treated as first-class entities within the DBMS. 
It enables the DBMS to jointly optimize relational and AI operators and coordinate their execution.

\end{definition}
To realize the paradigm, we distill two core design principles:

\begin{itemize}
    \item \textit{Holistic \aidb Co-Optimization}: \aidb workloads are iterative and concurrent, making decisions about relational processing and AI execution interdependent.
    The engine should therefore jointly optimize operator ordering, model placement, batching, and resource allocation across planning and execution.
    \item \textit{Unified \aidb Cache Management}: \aidb workloads offer reuse opportunities across executions. 
    The engine should therefore manage relational intermediates and AI artifacts through a unified cache abstraction to avoid redundant computation, data retrieval, model loading, and per-request initialization.
\end{itemize}

\begin{lstlisting}[
  float,
  caption={A representative ad-CTR \aidb query.},
  label={lst:query},
  basicstyle=\ttfamily\scriptsize,
  columns=fullflexible,
  breaklines=true,
  breakatwhitespace=true,
  lineskip=-1pt,
  numbers=none,
  belowcaptionskip=0.2em,
  belowskip=-3em,
]
/* Build point-in-time features for the 3-day horizon. */
@WITH@ pit_features @AS@ (
  @SELECT@ a.ad_id, c.ts, a.category_id, l.ctr_3d, agg(e.*) AS features
  @FROM@ candidate_ads a @JOIN@ cutoffs c @ON@ c.horizon = '3-day'
       @JOIN@ ad_events e @ON@ e.ad_id = a.ad_id @AND@ e.time <= c.ts
       @JOIN@ ctr_labels l @ON@ l.ad_id = a.ad_id @AND@ l.ts = c.ts
  @GROUP BY@ a.ad_id, c.ts, a.category_id, l.ctr_3d)
  
/* Invoke PREDICT to infer 3-day CTR for candidate ads. */
@SELECT@ ad_id, ts, nr_pred AS ctr_3d
@FROM@ (~PREDICT VALUE OF~ ctr_3d @FROM@ pit_features ~TRAIN tabpfn ON *~) p
@WHERE@ is_candidate_ad = true;
\end{lstlisting}

\section{Key Design Challenges}
\label{sec:challenge}

\subsection{Holistic \aidb Co-Optimization}
\label{sec:challenge-optimizer}

\subsubsection{Joint Optimization}
Given an \aidb query, the optimizer first performs semantics-preserving rewrites to obtain a simplified logical plan, and then maps it to a physical plan by jointly reasoning over DB and AI operators under declarative constraints.
An ideal design should provide the following capabilities.
\begin{itemize}
    \item \textit{Constrained optimization} enumerates and selects physical implementations for both DB and AI operators to optimize an end-to-end objective (e.g., latency/cost) subject to explicit quality constraints (e.g., accuracy/freshness bounds), or vice versa~\cite{LiSSF0BK23}.
    For DB operators (e.g., $\texttt{Scan}$, $\texttt{Join}$), it chooses join/aggregation algorithms, data partitioning/shuffling strategies, and parallelism policies.
    For AI operators (e.g., $\texttt{PREDICT}(\cdot)$), it chooses execution pipeline implementations and placement decisions.
    \item \textit{Cross-operator simplification} rewrites physical plans across relational and AI operator boundaries to eliminate work that does not affect the query result, thereby reducing unnecessary data access and model computation.
    For example, it may push filters and projections toward data sources to avoid reading or featurizing irrelevant inputs, or prune model components that are irrelevant to the query result during inference~\cite{ParkSBSIK22}.
    \item \textit{Cross-query optimization} reduces redundant work across concurrent queries by reusing cached relational intermediates and AI artifacts, while coordinating with the execution engine to materialize and retain results that may benefit future queries~\cite{MoCWCB23}.
\end{itemize}

\paragraph{Key challenges}
\textit{AI-specific rewrite space.}
\aidb queries require semantics-preserving AI-specific rewrites to eliminate redundant model computation.
For example, AISQL~\cite{abs-2511-07663} rewrites certain semantic joins from pairwise matching into multi-label classification when explicit preconditions are satisfied.
Optimizing such queries also requires determining the application order of mixed DB and AI rewrite rules.
However, the number of possible rewrite sequences grows combinatorially, and the effectiveness of each sequence is highly query-dependent~\cite{ZhouLCF21}.
Supporting both AI-specific transformations and efficient exploration of mixed DB--AI rewrite sequences within a unified compiler therefore remains an open challenge.

\textit{Composable AI-operator physicalization.}
The optimizer must consider a composable set of physical implementations for AI operators that integrate with relational execution while exposing choices in model placement, residency, and data movement.
For example, colocating AI execution with partitioned data avoids cross-node transfers, whereas routing data to a GPU with a resident model avoids model-loading and warm-up overhead.
Existing systems typically make placement and residency decisions outside the optimizer's search space, leading to suboptimal physical plans~\cite{abs-2511-07663,kumarasinghe2026ipdb,DorbaniYLM25,LiSSF0BK23}.

\textit{Accurate cost-quality modeling.}
AI operator latency and output quality are state- and data-dependent, which goes beyond classical DB cost models.
While discriminative models (e.g., gradient-boosted trees) can often be modeled using relatively stable and profile-based estimates~\cite{LiSSF0BK23}, generative models (e.g., LLMs) exhibit highly variable costs and quality: execution cost depends on token counts and cache state, while output quality varies with prompts, retrieved context, and decoding choices.
Consequently, coarse offline estimates may mislead the optimizer, yielding physical plans that incur unnecessary cost or fail to satisfy quality constraints~\cite{LiSSF0BK23,PanL25,abs-2503-23863}.

\textit{Safe cross-operator simplification.}
Cross-operator rewrites, such as projection pushdown, can reduce computation for traditional tree-based models~\cite{ParkSBSIK22}.
Extending them to foundation models remains challenging: complex featurization and execution pipelines, and high-dimensional learned representations obscure which computations can be safely pruned without changing query results.

\begin{figure}[t]
  \centering
  \includegraphics[width=\linewidth]{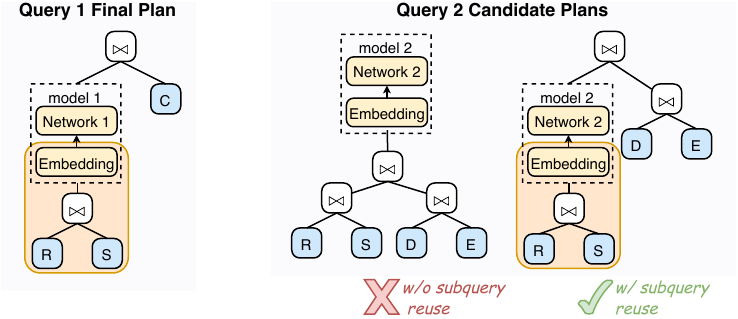}
  \caption{A motivating example for cross-query optimization.}
  \label{fig:motivation_opt}
\end{figure}

\textit{Cache-aware cross-query reuse.}
Concurrent \aidb queries often share data preparation, features, embeddings, or model invocations.
The optimizer should treat cached subplans as materialized intermediates and select plans that maximize safe reuse without violating data or model-version consistency.
As shown in Figure~\ref{fig:motivation_opt}, reusing the cached result of \texttt{HashJoin}(\texttt{R},\texttt{S}) and cached embeddings avoids redundant relational and AI computation.

\paragraph{Practical realization}
To realize these capabilities, AI operators can be exposed to the optimizer via a contract that specifies input dependencies, model versions, available physical implementations, resource demands, cost--quality profiles, and cacheable outputs.
The optimizer can then place AI and relational operators in a shared search space and perform cache-aware, bounded-objective planning over end-to-end latency, resource usage, and output quality.

\subsubsection{Self-Adaptive \aidb Co-Execution}
\label{sec:challenge-execution}
At runtime, the executor runs optimizer-selected physical plans and coordinates DB and AI operators under resource contention.
Specifically, it should
(i) reconcile stream-oriented DB processing with batch-oriented AI execution,
(ii) dynamically adjust batching, placement, and backpressure under concurrent load to balance per-query latency, aggregate throughput, resource efficiency, and performance isolation, and
(iii) preserve snapshot- and model-version-consistent semantics across batching, retries, migration, and model updates.
In particular, batches must not cross snapshot/model-version boundaries, retries and migration must preserve input lineage without exposing partial outputs, and model updates must become visible atomically.

\paragraph{Key challenges}
\textit{Consistency-preserving batching.}
AI operators often rely on batching for efficiency~\cite{XiaoRLZHLFLJ20}, but batching may reorder or group requests associated with different data snapshots or model versions, risking inconsistent query results.
Existing systems often lack database-style execution semantics for preserving such consistency during batching~\cite{abs-2509-02121,abs-2511-07663}.

\textit{Heterogeneous AI$\times$DB runtime scheduling.}
DB and AI operators have different runtime behavior and resource profiles.
Discriminative models often have relatively stable latency and resource usage, whereas generative models incur token-dependent costs that vary with prompt length, decoding, retrieved context, and caching state.
On shared accelerators, bursty arrivals or long-running generations can monopolize GPU memory/compute and amplify tail latency.
The runtime must therefore coordinate batching, placement, and backpressure across these heterogeneous operators.

\textit{Fine-grained failure recovery.}
Failures such as accelerator out-of-memory errors, kernel or runtime failures, and external model timeouts can interrupt execution after substantial partial progress.
Fine-grained recovery requires tracking input partitions, data snapshots, model versions, and reusable intermediates, allowing failed work to be retried or migrated without exposing partial outputs or recomputing valid work.
Most existing systems instead rely on coarse-grained, whole-request retries and lack database-style recovery that preserves snapshot and model-version consistency while reusing valid intermediates~\cite{kumarasinghe2026ipdb,abs-2511-07663}.

\paragraph{Practical realization}
To realize these capabilities, AI operators can run as stateful execution nodes that consume relational streams, form micro-batches, and attach partition, snapshot, and model-version metadata to each batch.

\subsection{Unified \aidb Cache Management}
\label{sec:challenge-cache}
\subsubsection{Unified Cache Abstraction}
A unified cache treats relational intermediates and AI artifacts (e.g., feature tables, embeddings, and model states) as first-class cacheable objects.
Unlike conventional DB caches of pages and blocks, these objects have heterogeneous structures, lifetimes, placement constraints, and validity dependencies.
A unified abstraction offers two benefits:
(i) correct cross-boundary reuse, by tracking data-snapshot and model-version dependencies in a common versioned key space across iterations and concurrent executions;
and (ii) global, plan-aware allocation, by giving the optimizer and runtime a shared view of placement, eviction, and reuse across memory tiers.

\paragraph{Key challenges}
\textit{Heterogeneous cache objects.}
The cache must represent heterogeneous objects (e.g., relational intermediates, embeddings, KV-cache blocks, model states, and optimizer states), whose different structures, granularities, lifetimes, and placement constraints complicate cache keys and equivalence rules.

\textit{Version-aware validity.}
A cached AI artifact can be safely reused only when its dependencies on data snapshots, model versions, prompts, feature definitions, and preprocessing pipelines remain valid.
The cache must track these dependencies precisely without imposing excessive lookup and invalidation overhead.

\subsubsection{\aidb Caching Policy}
\aidb caching policies decide which objects to materialize, where to place them (e.g., GPU/host memory or secondary storage), and when to reuse, evict, or refresh them.

\paragraph{Key challenges}
\textit{Workload-adaptive eviction.}
Traditional policies (e.g., LRU and FIFO) are ill-suited to \aidb workloads, whose reuse patterns are shaped by agent iterations, concurrent queries, model invocations, and plan-level dependencies.
The cache manager should therefore estimate reuse utility from runtime signals such as access frequency, reuse distance, and execution-plan dependencies.

\textit{Model-centric state management.}
AI operators generate states such as activations, gradients, optimizer states, embeddings, and KV-cache blocks.
These objects can dominate memory usage and differ from relational pages in their lifetimes, recomputation costs, and reuse patterns, requiring model-aware caching policies.

\textit{Multi-tier placement under concurrency.}
Relational intermediates and AI artifacts compete for capacity across GPU/host memory and secondary storage.
Under concurrent \aidb workloads, redundant transfers and uncoordinated placement can cause resource underutilization and increase tail latency.
The cache manager must therefore coordinate placement and eviction across memory tiers.

\paragraph{Practical realization}
Placement, eviction, and prefetch can be formulated as a learnable policy over a multi-tier memory hierarchy (GPU/host memory, secondary storage).
The policy can use runtime telemetry (e.g., access frequency, reuse distance, batching windows, GPU memory pressure, and tail-latency feedback) to decide which relational intermediates or AI artifacts to keep, move, or evict.

\section{\aidb architecture envisioning}
\label{sec:engine}

Guided by the envisioned design principles in Section~\ref{sec:vision} and the challenges identified in Section~\ref{sec:challenge}, we design and implement a proof-of-concept prototype, \sysname{}, for \aidb workloads.
\sysname{} runs on a shared disaggregated CPU/GPU substrate and consists of three main components: a holistic query optimizer, a self-driving execution engine, and a multi-tier cache manager.

\paragraph{Unified Declarative Interface}
\sysname{} provides two classes of SQL extensions: (i) model statements for managing models, and (ii) \texttt{PREDICT} statements for invoking models within analytical queries. 
As shown in Listing~\ref{lst:query}, \texttt{PREDICT VALUE OF} specifies the prediction target and consumes the relational input \texttt{pit\_features}.
The \texttt{TRAIN * ON *} clause specifies the model and feature columns, and the outer SQL query projects and filters the predicted values.

\begin{figure}[t]
  \centering
  \includegraphics[width=\linewidth]{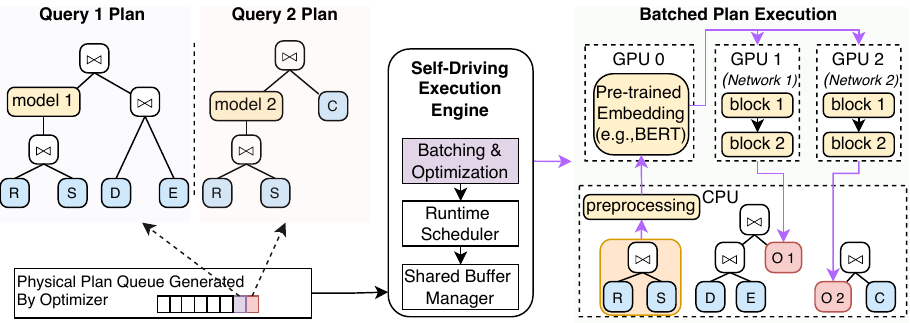}
  \caption{CSE-optimized execution graph in the execution engine.}
  \label{fig:engine_cse}
\end{figure}
\begin{figure}[t]
  \centering
  \includegraphics[width=\linewidth]{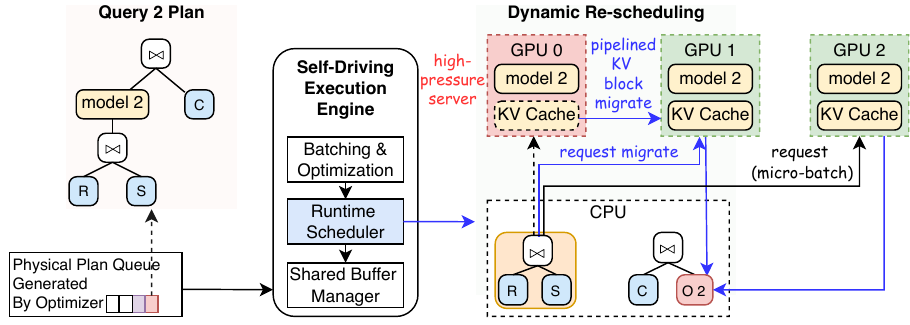}
  \caption{Dynamic rescheduling in the execution engine.}
  \label{fig:engine_schedule}
\end{figure}

\begin{figure}[t]
  \centering
  \includegraphics[width=\linewidth]{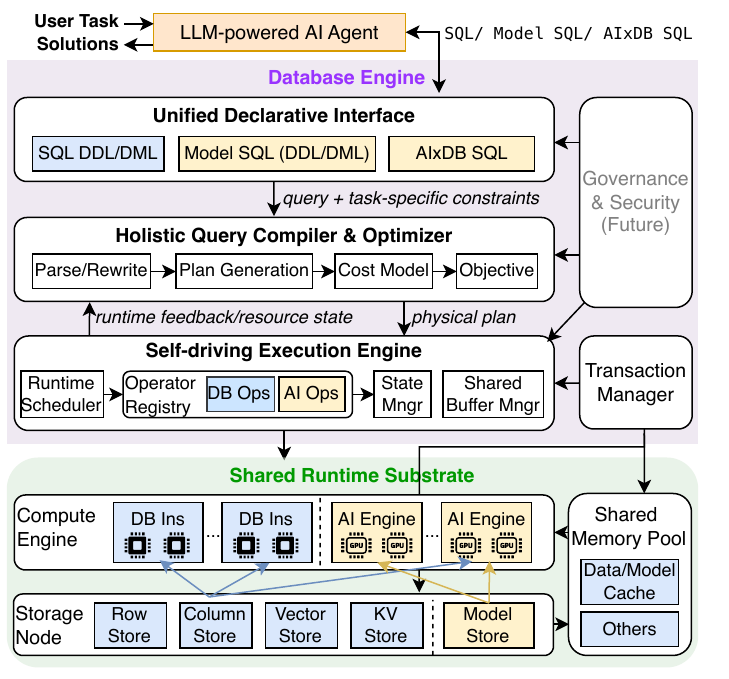}
  \caption{Envisioned \aidb architecture.}
  \label{fig:arch}
\end{figure}

\paragraph{Holistic Query Optimizer}
The compiler parses \texttt{PREDICT} queries into a logical plan.
Figure~\ref{fig:arch} depicts this stage as parse/rewrite, plan generation, cost modeling, and objective-driven optimization over constraints.
The optimizer then converts the logical plan into a physical plan by searching a joint space that includes 
(i) operator-ordering choices across DB and AI operators, and
(ii) physical implementation choices for both DB and AI operators.
In our prototype, this joint search is instantiated by AI-operator plan reordering (e.g., pushing selective predicates into the input subquery of \texttt{PREDICT} so the AI operator runs only on candidate rows) and cache-aware planning (materializing reusable subplans and reusing cached intermediates as building blocks for subsequent queries).

\paragraph{Self-driving Execution Engine}
\sysname{}'s executor runs physical plans with micro-batching and scheduling.
As shown in Figure~\ref{fig:arch}, the operator registry exposes DB and AI operator implementations to the scheduler, the state manager tracks execution state, and the shared buffer manager stores reusable artifacts.

To share work across queries, the executor identifies identical subplans and merges them into an execution graph, as illustrated in Figure~\ref{fig:engine_cse}.
Each shared subplan is executed once, and its results are routed to multiple downstream operators.
This cross-plan common subexpression elimination (CSE) reduces redundant I/O and computation while alleviating contention on shared resources.

The prototype also supports distributed routing of prediction tasks.
For example, the input to a \texttt{PREDICT} operator is partitioned into micro-batches and dispatched across multiple model instances.
The architecture further envisions dynamic rescheduling that migrates in-flight tasks and their execution states away from overloaded instances, as illustrated in Figure~\ref{fig:engine_schedule}.
This would support load balancing, performance isolation, and resource defragmentation.

\paragraph{Execution Semantics}
\sysname{} preserves snapshot and model-version consistency throughout
execution: each \aidb query reads from one MVCC snapshot, each
\texttt{PREDICT} uses a pinned model version, and batches never mix snapshots
or model versions. Results become visible only upon query completion, and model
updates are published atomically at commit.

\paragraph{Multi-Tier Cache Management}
\sysname{} materializes reusable subplan results, such as point-in-time feature tables shared across \aidb queries, through the shared buffer manager.
The optimizer treats these results as cached intermediates during cache-aware planning, avoiding repeated feature construction.
As shown in Figure~\ref{fig:arch}, the broader architecture generalizes this mechanism to a shared memory pool and unified data/model caches.

\paragraph{Workflow}
Figure~\ref{fig:arch} outlines the architecture; we explain it through the life cycle of an \aidb query.
An LLM-powered AI agent may translate a user task into SQL, model SQL, or \aidb SQL submitted through the unified declarative interface.
The query compiler translates the submitted query into a logical plan, and the holistic optimizer maps it to a physical plan by jointly considering relational processing, AI execution, and data/model movement.
The \aidb execution engine runs the plan with runtime scheduling and state management, returns query results, and materializes reusable artifacts.
A shared runtime substrate manages row/column storage, vector/KV stores, and model storage, while database fundamentals such as transaction management provide correctness and control.
Finally, execution statistics and states are fed back to the optimizer for adaptive optimization under dynamic workloads.

\section{Evaluation}
\label{sec:evaluation}
\subsection{Experimental Setup}
We run all the experiments on a server with 2 Intel Xeon Silver 4214R CPUs (24 cores/48 threads), 128GB RAM, and 8 NVIDIA RTX 3090 GPUs, with Ubuntu 22.04 Docker containers with CUDA 11.8.0.
We implement~\sysname{} in NeurDB~\cite{neurdb1, neurdb2}, an open source AI-powered database~\cite{neurdbgit} based on PostgreSQL v16.3.

\begin{figure*}[t]
  \centering
  \begin{subfigure}[t]{0.19\textwidth}
    \centering
    \includegraphics[width=\linewidth]{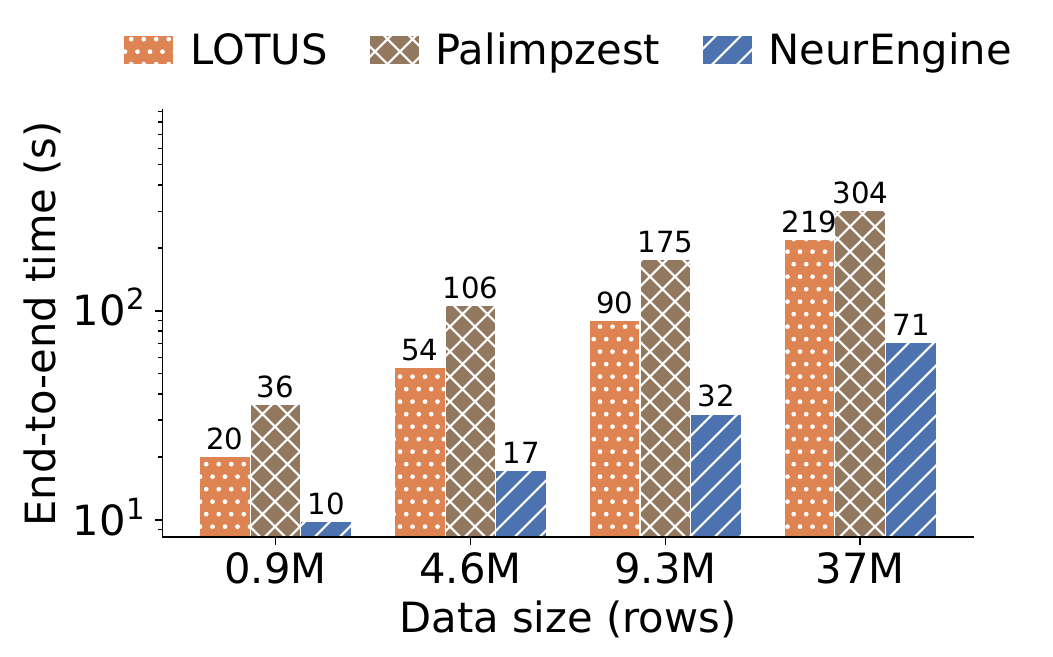}
    \caption{Data scalability}
    \label{fig:query:baseline}
  \end{subfigure}\hfill
  \begin{subfigure}[t]{0.19\textwidth}
    \centering
    \includegraphics[width=\linewidth]{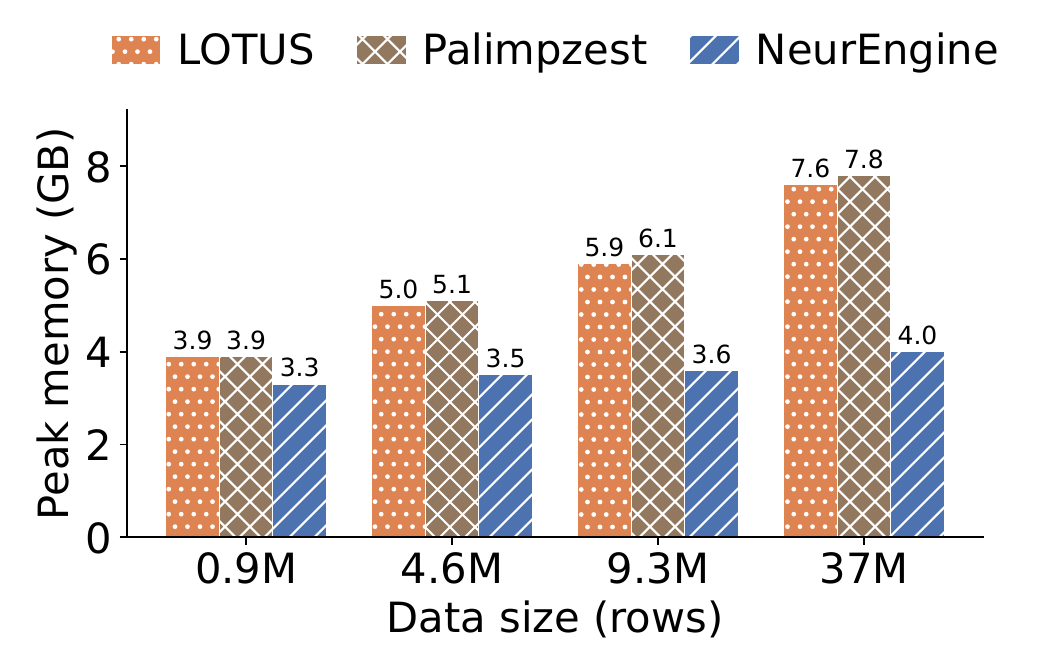}
    \caption{Memory footprint}
    \label{fig:query:memory}
  \end{subfigure}\hfill
  \begin{subfigure}[t]{0.19\textwidth}
    \centering
    \includegraphics[width=\linewidth]{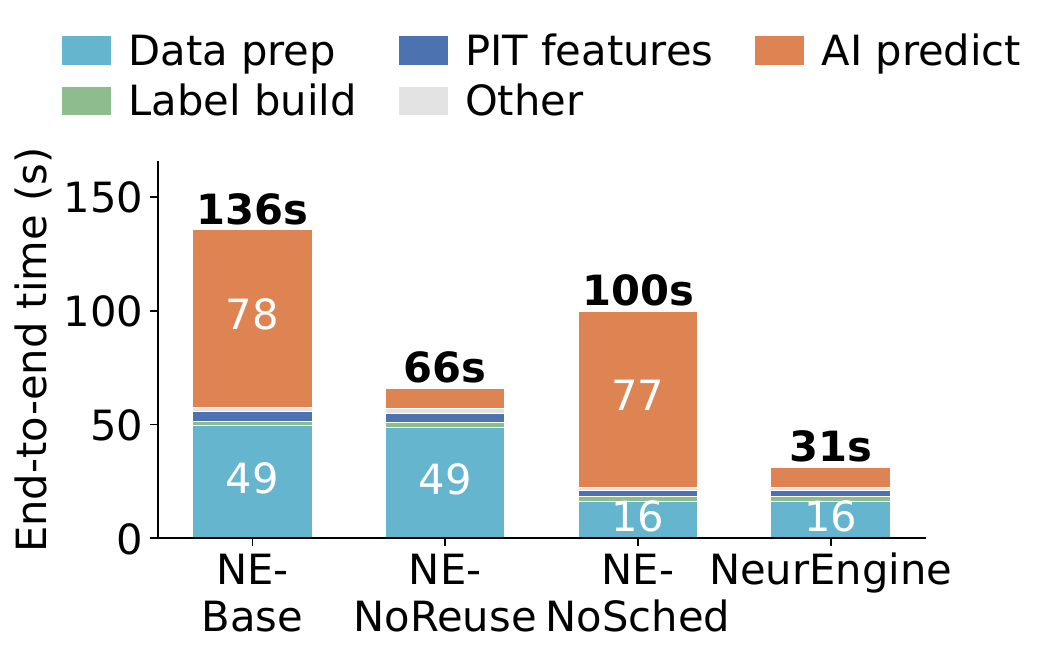}
    \caption{Execution breakdown}
    \label{fig:query:ablation}
  \end{subfigure}\hfill
  \begin{subfigure}[t]{0.19\textwidth}
    \centering
    \includegraphics[width=\linewidth]{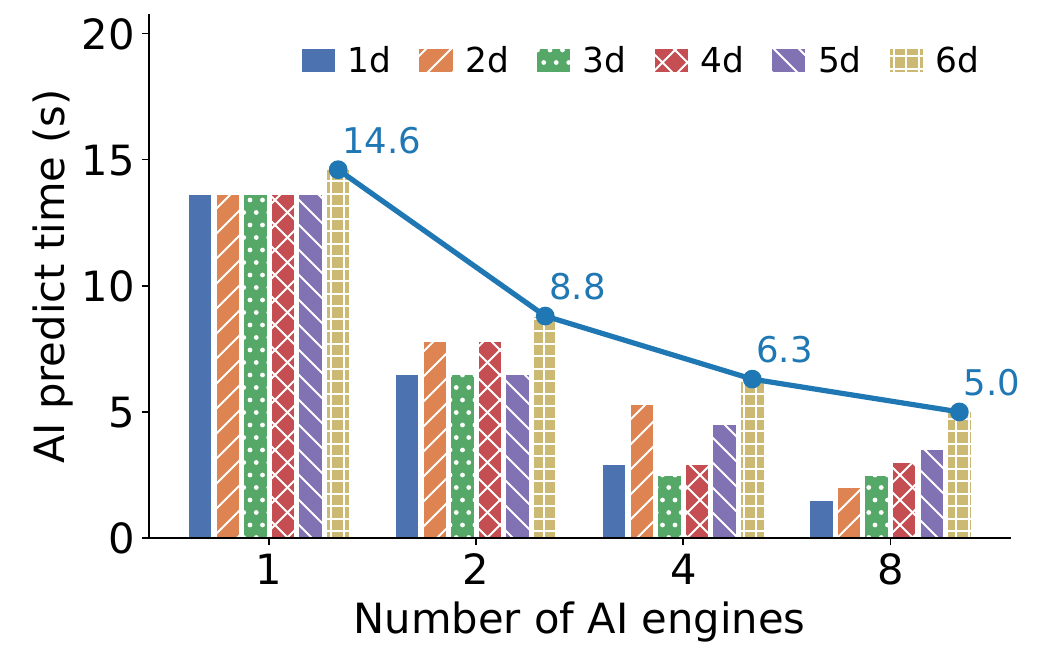}
    \caption{AI-engine scalability}
    \label{fig:query:scalability}
  \end{subfigure}\hfill
  \begin{subfigure}[t]{0.19\textwidth}
    \centering
    \includegraphics[width=\linewidth]{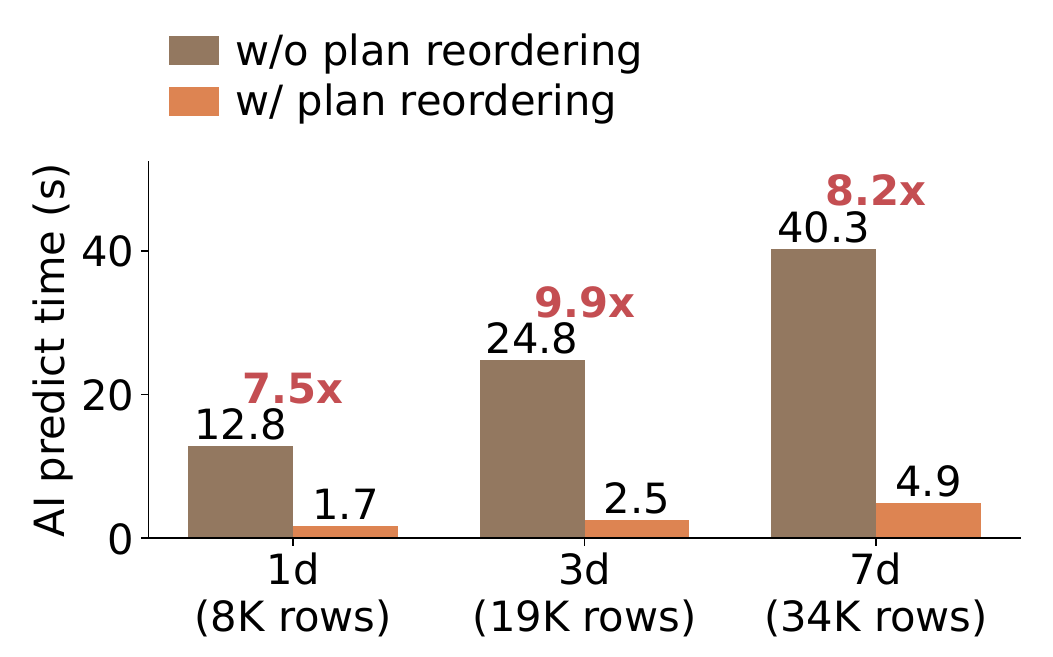}
    \caption{Plan reordering}
    \label{fig:query:scheduling}
  \end{subfigure}
  \caption{Evaluation of \sysname{}.}
  \label{fig:result_revision}
\end{figure*}

\paragraph{Workload}
We evaluate the ad click-through-rate prediction task introduced in Section~\ref{sec:vision} on the multi-table RelBench Avito dataset~\cite{relbench}, which spans six tables.
The AI agent translates the natural-language task into an \aidb workload: a task graph with three \aidb queries for the 1-, 3-, and 7-day CTR horizons, followed by a cross-horizon aggregation task.
Each prediction query builds labels and point-in-time features in SQL and invokes \texttt{PREDICT} with TabPFN~\cite{TabPFN}; the aggregation task joins the prediction results into the ad list.
To vary the data volume, we derive $0.1\times$/$0.5\times$/$4\times$ variants of the database by sampling or replicating the interaction events.

\paragraph{Baselines}
We compare with LOTUS~\cite{patel2025semantic} and Palimpzest~\cite{palimpzestCIDR}, two recent AI-powered analytics systems.
Neither provides a tabular prediction operator over multi-table relational data within an RDBMS, so our baselines use the closest supported execution path: export the relational tables to Parquet, build labels and point-in-time features with a shared pandas pipeline, and invoke TabPFN as opaque user code.
For Palimpzest, each \aidb query is expressed as a pipeline over a \texttt{MemoryDataset}, followed by \texttt{filter} and \texttt{map(UDF)} operators. 
For fairness, both baselines reuse \sysname{}'s type-aware preprocessing and TabPFN configuration, so all systems run the same algorithm over identical features and labels.

\subsection{Performance Analysis}
\textbf{Data Scalability.} 
We first compare the end-to-end time of the workload across the four data scales.
As shown in Figure~\ref{fig:query:baseline}, \sysname{} is the fastest at every scale, and the gap widens with the data volume: at 37M rows it completes in 71s versus 219s for LOTUS ($3.1\times$) and 304s for Palimpzest ($4.3\times$).
Since all systems execute the same algorithms, the gap comes from orchestration: the baselines must first export the growing base tables out of the database (185s at 37M, dominating their totals), whereas \sysname{} pushes the entire pipeline, including the AI operator, to where the data resides.

\noindent
\textbf{Memory Footprint.}
As shown in Figure~\ref{fig:query:memory}, the DataFrame-based baselines materialize all exported tables in memory, so their footprints grow with the data volume, up to 7.6GB (LOTUS) and 7.8GB (Palimpzest) at the largest data size.
\sysname{} stays nearly flat (3.3GB to 4.0GB) under a $40\times$ data increase: the model runtime is constant, and the relational side streams data through a bounded buffer pool instead of loading tables into memory, which is essential for scaling to datasets beyond RAM.

\noindent
\textbf{Execution Breakdown.}
Figure~\ref{fig:query:ablation} dissects the end-to-end time at the base scale by disabling the two mechanisms that instantiate our design principles: unified \aidb cache management, implemented as cross-task reuse of shared data-prep artifacts (point-in-time feature cache), and holistic \aidb co-optimization, implemented as AI-operator plan reordering.
Starting from the naive configuration (NE-Base, 136s), cache reuse alone cuts data preparation from 49s to 16s (NE-NoSched), plan reordering alone cuts AI prediction from 78s to 14s (NE-NoReuse), and the full co-design completes in 31s, a $4.4\times$ overall improvement.
The two mechanisms target different parts of the pipeline, and their benefits complement each other.

\noindent
\textbf{Scalability with AI Engines.}
We next vary the number of AI engines available for executing prediction tasks, extending the workload to six horizon-specific \aidb queries issued concurrently.
As shown in Figure~\ref{fig:query:scalability}, \sysname{} routes each query to an available AI engine and reduces the AI-prediction wall time from 14.6s with one engine to 5.0s with eight ($2.9\times$).
The residual time is dominated by the 6-day horizon query, whose larger data volume makes it the slowest prediction task.

\noindent
\textbf{Plan Reordering.}
Finally, Figure~\ref{fig:query:scheduling} isolates the effect of holistic \aidb co-optimization through plan reordering.
The candidate predicate selects only ${\sim}14\%$ of each task table.
Without reordering, \texttt{PREDICT} consumes the full task table and runs TabPFN over every row (12.8s--40.3s per horizon).
With cost-based reordering, the planner pushes the candidate predicate into the input subquery of \texttt{PREDICT}, so TabPFN runs only on candidate rows, reducing per-query latency to 1.7s--4.9s ($7.5\times$--$9.9\times$).

\section{Related Work}
\label{sec:related_work}
\noindent
\textbf{Query-centric AI analytics systems} optimize AI operators with data processing.
LOTUS~\cite{patel2025semantic} provides semantic operators over DataFrames with operator-level statistical accuracy guarantees, while Palimpzest~\cite{palimpzestCIDR} optimizes declarative AI analytics across cost, latency, and quality.
Within DBMSs, iPDB~\cite{kumarasinghe2026ipdb}, FlockMTL~\cite{DorbaniYLM25}, and AISQL~\cite{abs-2511-07663} integrate AI inference into SQL and optimize its execution.
Raven~\cite{ParkSBSIK22,KaranasosIPSPPX20} jointly optimizes data preparation and ML inference.
Li et al.~\cite{LiSSF0BK23} optimize model assignment and predicate ordering under constraints.
MASQ~\cite{BuonoPSI021} and TranSQL+~\cite{SunGWH25,SunL025} compile ML pipelines and LLM inference, respectively, into SQL.
ZIP~\cite{LinM23} integrates lazy imputation into query processing.

\sysname{} provides DBMS-native orchestration that jointly coordinates query planning, runtime scheduling, and shared artifact reuse across iterative and concurrent \aidb queries.

\noindent
\textbf{State-centric AI-in-DB systems} make AI state (e.g., features, model artifacts) first-class and optimize its computation, storage, 

\noindent
and reuse.
Feature systems~\cite{Liu0PZMSIKTC23,ZhouZQZCHL0WC25} focus on feature management and consistency.
InferDB~\cite{SalazarDiazGR24} accelerates inference by replacing full model execution with index-based lookup (e.g., indexed embeddings).
MorphingDB~\cite{WuXYWLGBZTXLC25} automates model management and cost-aware batch inference.
Other systems~\cite{evadb,mindsdb} provide SQL interfaces over external ML stacks~\cite{PaszkeGMLBCKLGA19,tensorflow}, and in-DB ML systems~\cite{madlib,gaussml,Jasny0KRB20,LiCCWZ17} push learning into the engine. 
nsDB advocates tighter neural/symbolic co-design~\cite{YuanTZZQ24}.
Commercial platforms \cite{amazon_sagemaker,ibm_db2_ai_for_zos,oracle_autonomous_database,sap_hana,azure_sql_database,google_cloud_ai_platform} provide managed ML/AI capabilities around databases as external services.

\sysname{} uses DB and AI artifacts as visible state, so the cache decisions 
can affect both planning and execution.

\noindent
\textbf{New database execution models} rethink plan granularity, scheduling, and hardware backends to support agentic workloads.
Deep Research~\cite{kraskacidr26} 
proposes an agentic analytics runtime with \emph{Context} objects.
TRA~\cite{YuanJZTBJ21} proposes a relational abstraction for distributed ML backends.
Tao~\cite{LiuLZT24} schedules the fine-grained, decomposed plans to meet SLOs under multi-tenancy.
Terabyte and related GPU database work study exchange/offloading bottlenecks on GPU clusters~\cite{WuCCIS25,CaoSIAK23,GandhiAFGZSCCI23}.
DBOS~\cite{dbos_cidr,dbos_codesign} further pushes this direction by turning OS state into tables and OS services into SQL.

Overall, these efforts tackle individual challenges.
\sysname{} is distinguished by database-native orchestration: AI operators such as \texttt{PREDICT}, together with their data/model artifacts, are treated as first-class entities inside the database optimizer and runtime.

\section{Conclusion and Future Directions}
\label{sec:conclusion}
In this paper, we envision \emph{database-native orchestration} as a foundation for efficient execution of \aidb workloads.
We formalize the \aidb workload model and derive two guiding principles, i.e., holistic \aidb co-optimization and unified \aidb cache management, to frame an open research agenda. 
We also introduce \sysname{} as a proof-of-concept prototype and report preliminary results to demonstrate the feasibility of this direction.

Beyond the two core principles, database-native orchestration also requires governable \aidb execution, which we leave as future directions.
For access control, AI operators turn data access into model access: queries may route sensitive tuples through featurization, inference, and caching, and may produce derived artifacts with leakage risks such as membership inference~\cite{ShokriSSS17}, attribute inference~\cite{LuoWXO21}, and embedding inversion~\cite{QinLMHXZ26}.
For isolation, \aidb workloads mix short updates with long-running, multi-stage AI tasks that share models and artifacts, calling for stage-aware isolation and forkable transactions for speculative agent execution.

We hope this work helps crystallize the research agenda and motivates the community to develop shared abstractions, benchmarks, and system foundations for building effective, resource-efficient, and governable \aidb database engines.


\bibliographystyle{ACM-Reference-Format}
\bibliography{references}

\end{document}